# Nanomechanical subsurface characterisation of cellulosic fibres

Julia Auernhammer[1], Markus Langhans[2], Jan-Lukas Schäfer[2], Tom Keil[3], Tobias Meckel[2], Markus Biesalski[2] and Robert W. Stark[1*]

[1] Institute of Materials Science, Physics of Surfaces, Technische Universität Darmstadt, Alarich-Weiss-Str. 16, 64287 Darmstadt, Germany

[2] Department of Chemistry, Macromolecular Chemistry and Paper Chemistry, Technische Universität Darmstadt, Alarich-Weiss-Str. 8, 64287 Darmstadt, Germany

[3] Institute of Materials Science, Physical Metallurgy, Technical University of Darmstadt, Alarich-Weiss-Str. 2, 64287 Darmstadt, Germany

*corresponding author
stark@pos.tu-darmstadt.de

ORCIDs:

| | |
|---|---|
| Julia Auernhammer | 0000-0002-9896-7353 |
| Markus Langhans | 0000-0002-1117-5645 |
| Jan-Lukas Schäfer | 0000-0002-6959-9467 |
| Tom Keil | 0000-0002-9145-4731 |
| Markus Biesalski | 0000-0001-6662-0673 |
| Robert W. Stark | 0000-0001-8678-8449 |

*Acknowledgements:*

The authors would like to thank the Deutsche Forschungsgemeinschaft under grant PAK 962 project numbers 405549611, 405422473, 405440040 and 52300548 for financial support. Furthermore, we thank Dr. Martin Dehnert from the Technische Universität Chemnitz for support with the analysis software.

# Abstract

The mechanical properties of single fibres are highly important in the paper production process to produce and adjust properties for the favoured fields of application. The description of mechanical properties is usually characterised via linearized assumptions and is not resolved locally or spatially in three dimensions. In tensile tests or nanoindentation experiments on cellulosic fibres, only one mechanical parameter, such as elastic modulus or hardness, is usually obtained. To obtain a more detailed mechanical picture of the fibre, it is crucial to determine mechanical properties in depth. To this end, we discuss an atomic force microscopy-based approach to examine the local stiffness as a function of indentation depth via static force-distance curves. This method has been applied to linter fibres (extracted from a finished paper sheet) as well as to natural raw cotton fibres to better understand the influence of the pulp treatment process in paper production on the mechanical properties. Both types of fibres were characterised in dry and wet conditions with respect to alterations in their mechanical properties. Subsurface imaging revealed which wall in the fibre structure protects the fibre against mechanical loading. Via a combined 3D display, a spatially resolved mechanical map of the fibre interior near the surface can be established. Additionally, we labelled fibres with carbohydrate binding modules tagged with fluorescent proteins to compare the AFM results with fluorescence confocal laser scanning microscopy imaging. Nanomechanical subsurface imaging is thus a tool to better understand the mechanical behaviour of cellulosic fibres, which have a complex, hierarchical structure.

# Introduction

Paper, as a high-tech material made from cellulose, has promising applications in areas such as electronics, sensor technology, microfluidics and medicine (Bump et al. 2015; Delaney et al. 2011; Gurnagul and Page 1989; Hayes and Feenstra 2003; Liana et al. 2012; Ruettiger et al. 2016). Cellulose is a natural material, is abundant and renewable, and is the most important raw material in the papermaking industry. During papermaking, the natural structure of fibres can be mechanically or chemically altered, particularly at the fibre surface. To better understand how these alterations may affect the mechanical stability or water uptake of the fibres, it is essential to investigate the mechanical fibre properties at and beneath the surface. To tailor paper to advanced applications, the impact of fibre or pulp treatment before the paper-making process on the structure and mechanical properties of cellulose fibres must be understood. In particular, the fibre surface is relevant in technology because its composition and roughness determine the fibre-fibre bond strength and because it is exposed to the environment on the paper sheet surface (Rennel 1969; Rohm et al. 2014; Vainio and Paulapuro 2007). The properties of the bulk, however, are of similar relevance because they determine, for example, how well fluids such as water can be absorbed and how the fibre swells in a humid environment. To better understand how cellulose fibres swell and thus change their mechanical stabilities and how this process is related to the pulp treatment process of the papermaking procedure, it is essential to characterise bulk and surface properties close to the interface between the fibre and the surrounding atmosphere.

Cellulosic fibres have a hierarchical structure. By forming larger networks of cellulose molecules, highly oriented linearized polymer chains can form a unit cell. This unit cell is surrounded by unordered areas. With the arrangement of oriented (crystalline) and unordered (amorphous) areas, fibril structures are developed. The arranged microfibrils have a diameter of a few nanometres and are a few micrometres in length. By aggregation of microfibrils, macrofibrils with a diameter of 60-400 nm and a length of a few millimetres are formed (H.P. Fink 1990). These fibrils, together with hemicellulose, lignin, pectin or waxes, form the cell walls of plant-based fibres. The formed fibril structures possess a hierarchical wall structure, as shown in figure 1. A cotton fibre has a central, hollow lumen. Around the lumen, a secondary wall with layers S3, S2 and S1 is formed. This wall is surrounded by the primary wall (P). The mature fibre is enclosed by the cuticle (C), a waxy protection layer, which is a few molecules thick. Before processing the fibre, the cuticle must be removed. As seen in figure 1, the cellulose fibrils in the P wall are arranged in a disordered network around the fibre axis. The wall thickness of P is 0.1-0.2 $\mu$m, and it contains pectin, hemicelluloses and cellulose microfibrils and small amounts of cutin/wax and proteins. During the paper production process, this wall is often milled off. For the fibre or pulp treatment process, it is essential to remove C and a small amount of P, as the fibres are pressed in a wet state to bond to each other and thus form a paper sheet. Because of the missing C layer and intact P wall, the water molecules can intrude into the fibre network and break the hydrogen bonds between the cellulose molecules, which leads to softening of the fibres, which are maintained through the insolubility of cellulose (Cabrera et al. 2011; Gumuskaya et al. 2003; John and Thomas 2008; Lindman et al. 2010). The thickness of the S1 layer is 0.1-0.2 $\mu$m, and it contains small portions of pectin and hemicelluloses and high portions of cellulose. The fibrils orient in a predominant direction and lie parallel at an angle of 20° (microfibril angle MFA) to the lumen. The most relevant layer for paper production is the S2 layer. A switch in the fibril direction is observable in the transition region from S2 to S1. The fibrils here lie side by side in a predominant direction oriented along the fibre axis. With a thickness of 1-5 $\mu$m, the S2 wall represents 90 % of the fibre mass and therefore determines the mechanical properties of the fibre. The S3 layer separates the lumen from S2 and is 0.1-0.2 $\mu$m thick. The angle of the fibrils is 45 ° to the fibre axis. The changing and spiralled arrangement of the fibrils results in high inherent robustness of the fibres (Mather and Wardman 2015; Ott 2017; Sctostak 2009).

To date, tensile tests have characterised the tensile strength or elastic modulus of cellulosic fibres. An experimental approach to determine the longitudinal elastic modulus has been described by (Page et al. 1977). Theoretical works, e.g., (Mark and Gillis 1973) and (Salmen and Deruvo 1985), established that the MFA in the S2 layer is the determining factor for the strength of the fibre. Soon, it was established that the MFA in the S2 layer is the factor that determines the strength of the fibre. A small MFA angle leads to a high longitudinal elastic modulus (Müssig 2010). In addition, (Barnett and Bonham 2004) found that the mechanical properties depend on the orientation of the microfibrils in the S2 layer. Likewise, (Spokevicius et al. 2007) determined that the more closely the microfibrils are longitudinally aligned with each other in the S2 layer, the more tensile force could be applied. However, the transverse elastic modulus depends, according to *Bergander and Salmen*, less on the S2 layer than on S1 and S3 (Bergander and Salmen 2000). Nanoindentation was introduced as a method to investigate the different layers in the fibre. (Wimmer et al. 1997) determined the Young's modulus and the hardness of wood fibres via nanoindentation. (Gindl and Schoberl 2004) expected in their nanoindentation experiments that the Young's modulus of the S2 layer should be higher than the modulus of the other layers.

Sensitive nanoindentation can be performed by atomic force microscopy (AFM), which even makes it possible to map mechanical properties on the surface (Fischer et al. 2014; Piantanida et al. 2005). A straightforward approach to obtain a picture of the landscape of local mechanical properties is "force-volume mapping". In this approach, the force is measured via the cantilever deflection, which leads to force-distance curves. (Roduit et al. 2009) extended this method by introducing the "stiffness tomography" method, where they evaluated static force-distance curves in segments to show the stiffness differences along the indentation path. Thus, it is possible to estimate the Young's modulus of the sample at a desired indentation spot for various indentation depths. Previously released studies have shown that this approach can be applied to soft materials such as cells (Stuhn et al. 2019), polymers (Dehnert and Magerle 2018), bacteria (Longo et al. 2013), graphene oxide (Dehnert et al. 2016) or collagen fibrils (Magerle et al. 2020).

In the following, we discuss how AFM-based nanoindentation can be used to probe the near-surface bulk of cellulose fibres under varying environmental conditions (relative humidity). Variations in the local nanomechanical properties could be established and related to recovery from the hydrated state.

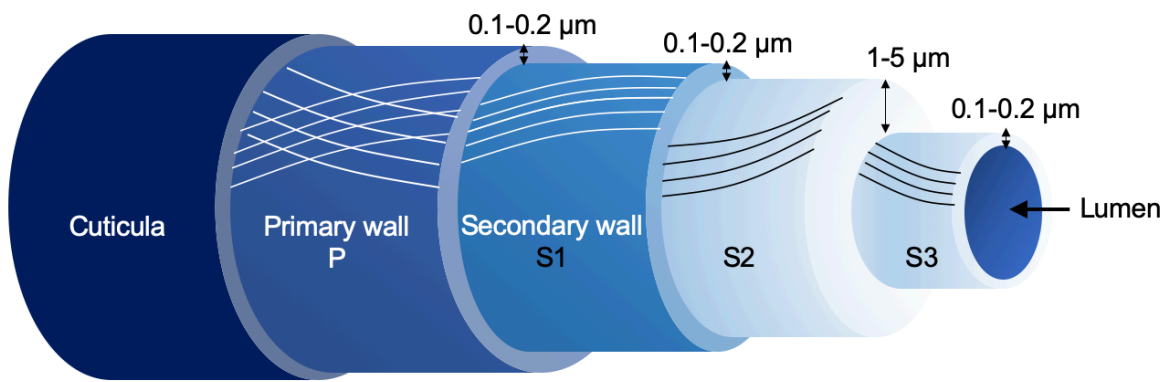

Figure 1: Schematic layered wall structure of a cotton fibre. The lines in the different walls represent the fibril arrangement inside the walls.

## Materials and Methods

<u>Fibres</u>

Processed cotton linter fibres were extracted manually from a linter paper sheet. The sheet was produced according to DIN 54358 and ISO 5269/2 (Rapid-Köthen process). As these fibres were already processed, they are referred to as PF in the following. Unprocessed cotton fibres were manually extracted from a dried natural cotton boll. These fibres were raw and unprocessed and will be referred to as UPFs. The fibres were fixed on both ends on glass substrates. For hydration, fibres were stored in deionised water and allowed to swell for 45 minutes. As shown in (Carstens et al. 2017; Hubbe et al. 2013), the water progressed into the cotton linter test stripes in seconds. Furthermore, it was discovered that the free swelling time for a whole pulp was 70 minutes (Olejnik 2012). However, (Mantanis et al. 1995) recorded the swelling of cellulose single fibres in water and showed that equilibrium was reached at 45 minutes. We performed the measurements in the wet state according to (Auernhammer et al. 2021). Additionally, the water drop for swelling was removed before the measurement, preventing the fibre from swelling further.
Overall, 9 spots of every fibre type and condition were investigated.

<u>Atomic Force Microscopy</u>

A NanoWizard II atomic force microscope (JPK InstrumentsAG, Berlin, Germany) was used to record maps of force–distance curves of the PF and UPF. Two types of cantilevers were used. The first type was the RTESPA-525 (Bruker, Santa Barbara, USA) with a high spring constant (HSC) of 162 N/m, a 15° opening angle only at the tip end and a radius of 30 nm. The second type was the ISC-125 C40-R (Team Nanotec, Villingen-Schwenningen, Germany) with a high aspect ratio (HAR), an opening angle of 10° for 3 $\mu$m, a radius of 10 nm and a spring constant of 30 N/m. The image size was 10 $\mu$m with 128x128 data points, a scan speed of 20 $\mu$m/s and a setpoint of 3000 nN.

<u>Force-Volume Mapping</u>

Performing force-distance curves allows not only investigation of the surface nanomechanics but also probing of the mechanical properties near the surface. To map sub-surface properties, i.e., to obtain a three-dimensional characterisation of the mechanical properties, the force-distance was analysed stepwise (30 nm) after the contact point $d_0$. Local slopes were interpolated stepwise to obtain estimates of the effective local elastic modulus as a function of depth $E_{lok}(z)$ (see figure 2). A force-distance curve extracted from the measurements is shown in figure S2.

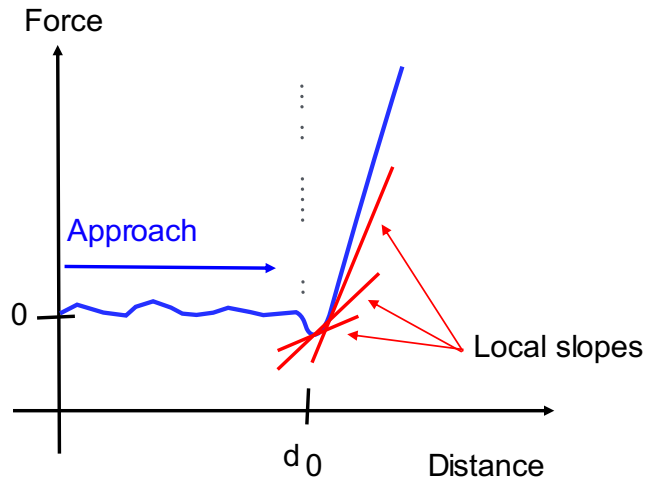

Figure 2: Schematic display of the stepwise analysis of a force-distance curve. Local slopes are marked in red. The indentation depth is calculated from the contact point $d_0$.

Before analysing the local slopes, the force-distance curves had to be processed. The noise was removed by a Savitzky-Goly filter. The slope of the baseline was corrected with a linear fit function to the initial flat region. The offset and the inclination of the curves were corrected. The contact point $d_0$ was identified as the first point in the repulsive regime (positive slope). More precisely, the first contact point was identified as the point where the force exceeded that of the preceding point by at least 10 times the standard deviation of the baseline signal. The position of the tip over the height $z$ was calculated from the deflection of the cantilever, which transformed the force-z-piezo data to a force-distance or force-indentation curve, respectively.

From a two-dimensional map of force-distance curves, various parameters were calculated. The topography was calculated from the contact points $d_0$. The penetration depth reached at the setpoint (maximum force) yielded the maximal indentation. To obtain mechanical information at intermediate depths, the data were interpolated in a stepwise manner. Thus, for the intermediate penetration depths, local slopes were calculated, which served as estimates of the effective elastic modulus assuming Hertzian contact mechanics (Hertz 1881). To obtain good spatial and depth resolution, the AFM cantilever was selected such that structures beneath the surface could be probed. This was accomplished via two different approaches: (i) by using hard cantilevers with a high spring constant (HSC) but conventional tips or (ii) by using softer cantilevers equipped with extremely sharp tips with a high aspect ratio (HAR). By using the HSC cantilever, a sufficiently large loading force can be applied that the tip probes the structure inside the fibres. In contrast, with the HAR cantilever, a large local pressure can be obtained at moderate loading forces. Both tip geometries are displayed in figure 3 with SE images. The data were processed by using MATLAB code. The two-dimensional images were generated by using Gwiddion (Necas and Klapetek 2012), and three-dimensional images were created by using Blender (Community 2018).

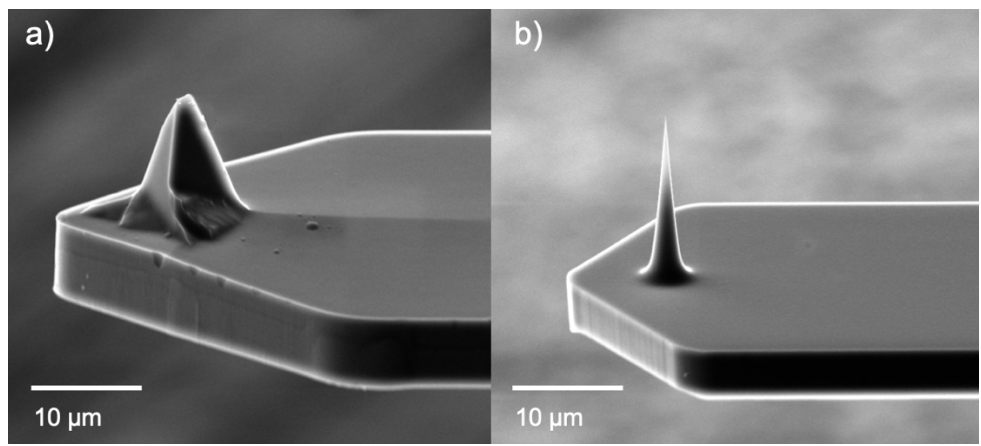

Figure 3: SE images of the tip geometry for a) the HSC cantilever and b) the HAR cantilever.

Scanning Electron Microscopy

The individual cantilever tip shapes were analysed by a scanning electron microscope (SEM) (MIRA3, TESCAN, Brno, Czech Republic) using secondary electron (SE) imaging mode at an acceleration voltage of 10 kV.

Preparation of Recombinant His- and Fluorescence-Tagged Carbohydrate-Binding Proteins (CBMs)

The DNA sequences of CBM3a (Blake et al. 2006; Fox et al. 2013; Venditto et al. 2016) ("semi-crystalline" cellulose), CBM1Cel6a (crystalline cellulose) (Blake et al. 2006; Fox et al. 2013) and CBM77 (Yaniv et al. 2013) (pectin) were taken from the CAZy database and linked DNA or protein databases (GenBank, UniProt, PDB). The evaluated sequences of *Clostridium thermocellum* (CBM3a; CCV01464.1, 4JO5), *Rumminococcus flavefaciens* (CBM77; WP_009983557, 5FU5) and *Trichoderma resii* (CBM1Cel6a; AAA34210.1, P07987) were then ordered as gene fragments (IDT, Coralville, USA) and cloned via Gibson assembly into a self-designed pET28 vector derivative. CBM-XFP-6xHis fusion proteins were expressed in *E. coli* BL21 cells by induction with IPTG. According to the protocols of (Blake et al. 2006) and "Polymer Probes And Methods" (Beauregard), induced cells were homogenised and lysed by the utilisation of EmulsiFlex-C3 from AVESTIN® in His-Trap Binding Buffer (20 mM NaH2PO4/0.5 M NaCl/pH 7.4 with NaOH). The desired His-tagged proteins were purified by immobilised metal affinity chromatography (IMAC) using a Ni-IDA column (Protino Ni-IDA 2000 packed column, Macherey-Nagel, Düren, Germany) and eluted with 200 mM imidazole. Elution fractions were pooled and concentrated using ultrafiltration unit filters (Vivaspin® 20; membrane 10,000 MWCO PES, Sartorius, Germany). During centrifugation, the imidazole buffer was exchanged into CBM storage buffer (50 mM Tris HCl; 20 mM NaCl; 5 mM CaCl2 x 2H2O; pH 7,4) via diafiltration.

3D Fluorescence CLSM Measurements

Confocal xyz series of CBM3a-mClover3-, CBM1Cel6a-mKOk- and CBM3a-smURFP-labelled fibres were recorded with a Leica TCS SP8 confocal system (Leica Microsystems GmbH, Mannheim, Germany) using an HyD detector with an HCX PL APO 63x NA 1.2 W CORR CS2 objective and the normal scanner system at 512 x 512 pixels in the 12-bit mode. Z-sections were set at a system-optimised value of 0.36 $\mu$m or a custom value of 0.2 $\mu$m per section. Sections were obtained using an appropriate laser for excitation and a small range of emission, 10 to 15 nm around the emission maximum (mClover3 ex. 488 nm, em. 505-525 nm; mKOk ex. 561 nm, em. 570-590 nm; smURFP ex.635 nm, em. 660-690 nm). Fluorescence channels were obtained in sequential frame detection mode to avoid cross talk.

# Results and discussion

From the force curve data, a surface topography map was calculated. In addition, maps of the local stiffness at various indentation depths were generated. The combined 3D representation of the topography with the corresponding local stiffness is shown in figure 4.

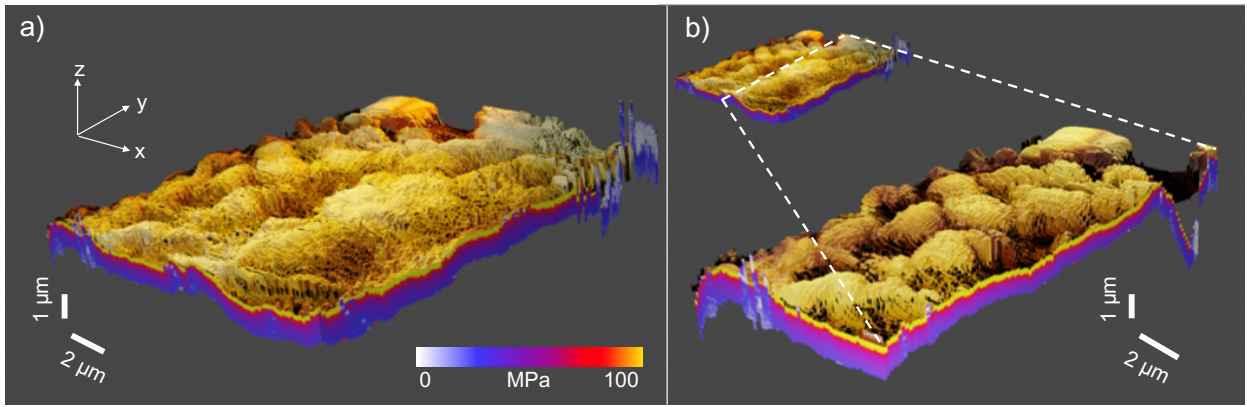

Figure 4: a) 3D representation of the surface profile of a PF. The colour in the z-direction encodes the local stiffness beneath the surface. b) A cross-sectional profile of $E_{lok}(z)$ in the xz- and yz-directions as indicated.

Figure 4a) shows an overview of the PF surface with its local stiffness maps. A cross section through the fibre with the corresponding local stiffness maps of $E_{lok}(z)$ is shown in the xz- and yz-directions in figure 4b). The parameter for the local stiffness as a function of the depth beneath the surface $E_{lok}(z)$ is meaningful in regions where the fibre topography could be mapped with sufficient resolution. "Sufficient resolution" here means a feature size of 12.8 pixels/$\mu$m.

In figure 5, an overview of cross sections of $E_{lok}(z)$ in the xz-direction in the PF and UPF mapped with the different cantilever types are shown.

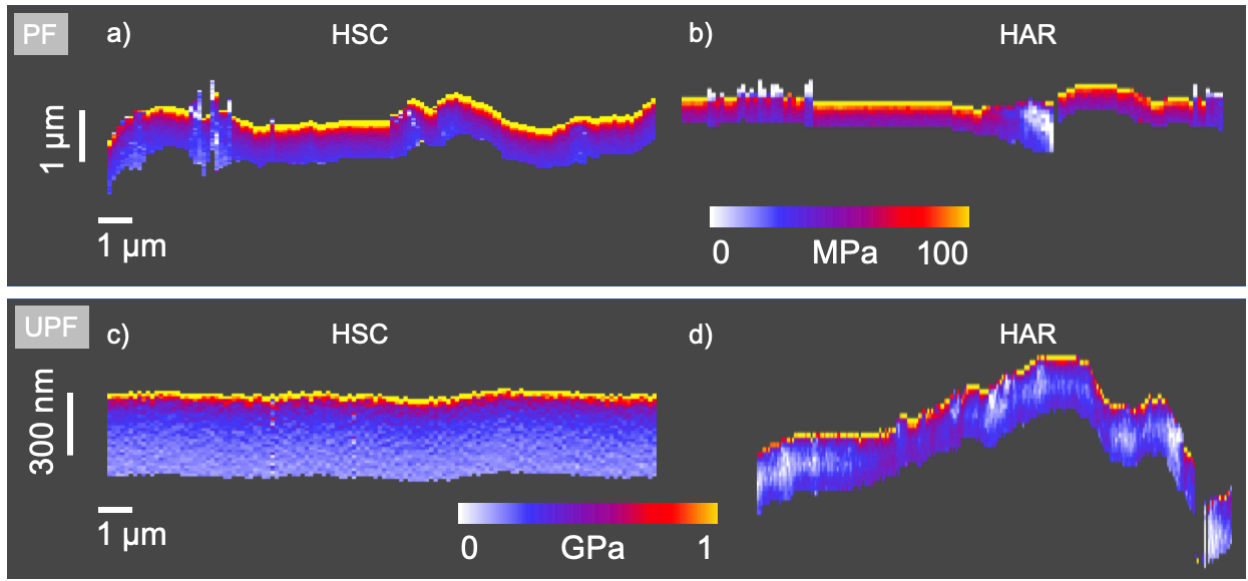

Figure 5: Depth profile of $E_{lok}(z)$ in the xz-direction of a PF mapped with an HSC cantilever in a) and an HAR cantilever in b) and of UPF mapped with an HSC cantilever in c) and an HAR cantilever in d).

In figure 5a), the depth profiles of $E_{lok}(z)$ in the xz-direction of a PF mapped with a HSC cantilever in a) and with a HAR cantilever in b) are shown. Both depth profiles exhibited a hard layer, shown in in yellow. The hard layer on the top exhibits a value of 95 ± 15 MPa. A sharp transition between a hard layer (yellow) and softer material (red and blue) can be identified. Thus, a stiffness gradient from the fibre surface to the softer interior was observable. $E_{lok}(z)$ was calculated for each data point, for a total of 128 profiles. All of them showed the same behaviour with the hard layer at the surface and sharp transition to the softer layer beneath. To verify the trend of $E_{lok}(z)$ in the xz- and yz-directions, further profiles are displayed in figure 4b), where the topography and the profiles of $E_{lok}(z)$ are shown for another cross section. Additionally, here, a hard layer (yellow) can be identified above the softer layers (red and blue). Thus, it is reasonable to assume that the PF is covered by a hard layer (or several hard layers that could not be resolved). Beneath this top layer, $E_{lok}(z)$ decreases with further indentation depth. Cross sections of the $E_{lok}(z)$ of PF are shown in figure 6b).

However, in the fibre or pulp treatment process, the P wall is usually milled off. Thus, it has to be excluded that the hard layer on top in the PF presents the S2 layer. To verify this, force-volume mapping was applied to the UPFs. The UPFs were directly extracted from a natural cotton boll and therefore raw and unprocessed, which means that all layers should be intact. In figure 5c), a depth profile of $E_{lok}(z)$ in the xz-direction mapped with an HSC cantilever and in d) mapped with a HAR cantilever are shown. From figure 5c) and d), it can be seen that the UPF possessed a hard layer (yellow) on the fibre surface with softer layers beneath (colour coded in red and blue). The $E_{lok}(z)$ in the xz-direction decayed with increasing indentation depth. Compared to the PF, the UPF exhibits a harder layer on the surface at 1 GPa. The indentation depth in all measurements was sufficient to reach the S2 layer depth, i.e., at least 300 nm. However, it was observed that the indentation depth in the UPFs was not as high as that in the PFs. Cross sections of the $E_{lok}(z)$ of UPF are shown in figure 6a).

We interpret the trend of the depth profile of UPF as follows: In UPF, all layers should be intact. Therefore, the hard layer (yellow) on the top most likely represents the $E_{lok}(z)$ of C with incrustations (eventually with a slight crosstalk caused by P). Layer C is a waxy layer containing cutin, waxes and cell wall polysaccharides, which is assumed to be harder than the layer beneath it, which is embossed by the fibril structure of cellulose. The slightly softer layer beneath (red) might directly represent the P wall. In comparison to the C layer, the $E_{lok}(z)$ in the P wall is reduced because it contains additives to waxes, such as lignin, minerals and cellulose fibrils. Compared to the S layers, which are predominantly constructed of cellulose fibrils, the P wall with waxes, lignin or minerals is therefore harder. At the indentation depth, where S1 and S2 should be present, $E_{lok}(z)$ decayed noticeably.

Regarding the question of what happens to the fibres when they are processed into paper fibres, we propose the following: In the PF, a maximum average value of $E_{lok}(z)$ 87 ± 16 MPa was measured (hard layer on the top). In UPF, this $E_{lok}(z)$ value is reached at an indentation depth of 235 ± 58 nm. As the depth of S2 should be 300-500 nm, this result suggests that in PF, C was completely removed during the paper production process. Additionally, the P and S1 layers were slightly milled off in PF, and the lignin content decreased. However, the data indicate that the P and S1 layers were not completely removed during paper production, as the $E_{lok}(z)$ of the S2 layer was not reached in the first 200 nm. Since the P wall has a thickness of 100-200 nm, as does the S1 layer, it can be assumed that both layers could still be partially intact.
In figure 6c), further cross sections of $E_{lok}(z)$ UPF (in red) and PF (in blue) are shown. The suggested corresponding wall layers are indicated with different colours in the graph (see C, P, S1,

S2). The C layer is highlighted in dark blue, the P wall in blue, the S1 layer in light blue and the S2 layer in grey. Figure 6c) represents the suggested $E_{lok}(z)$ trend inside the fibre. The cross sections of $E_{lok}(z)$ of the PF begin where the $E_{lok}(z)$ value of the UPF fibre is matched.

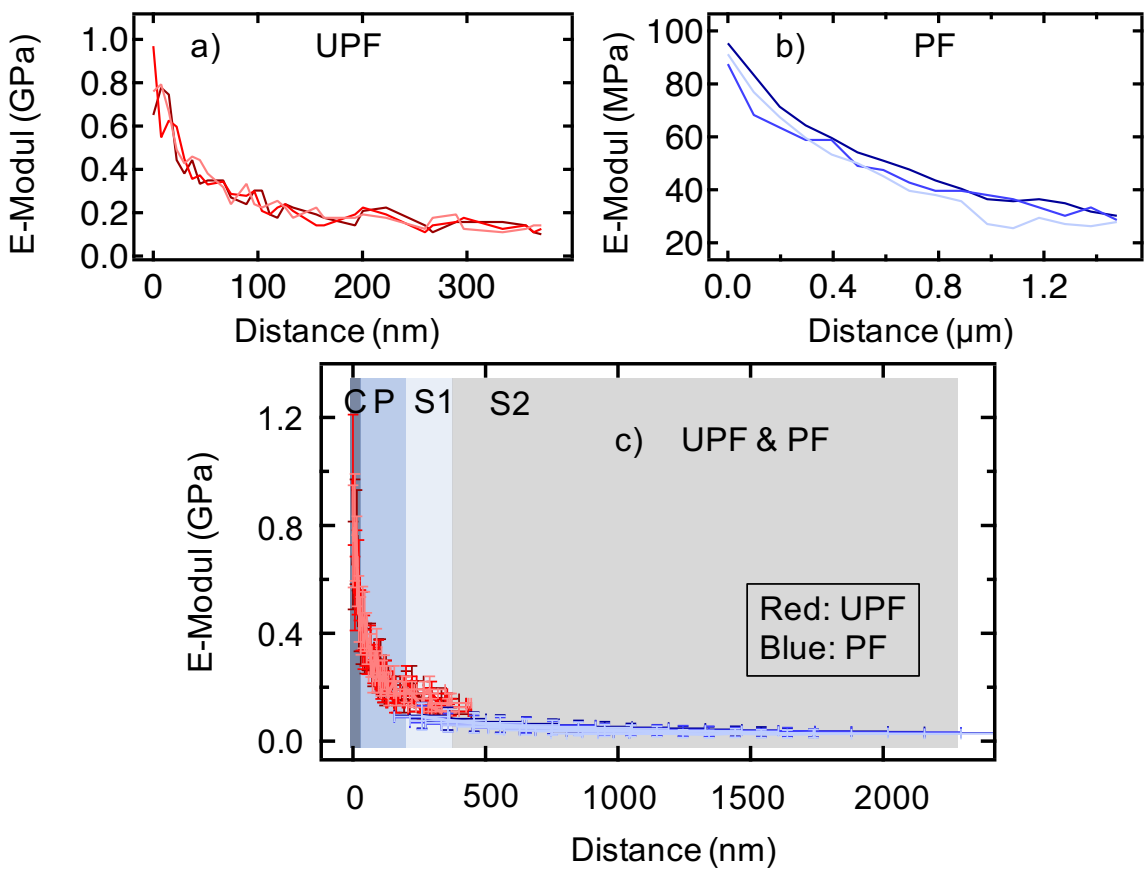

Figure 6: Depth profiles of this $E_{lok}(z)$ of a) a UPF, b) a PF and c) a combined graph of UPF and PF. In c), the UPF profiles are plotted in red, and the PF profiles are plotted in blue. The suggested corresponding wall layers are indicated with different colours. The C layer is highlighted in dark blue, the P wall in blue, the S1 layer in light blue and the S2 layer in grey. The error bars are ±30 % of the values.

To verify our suggested wall structure, we labelled UPF and PF with fluorescence protein-tagged CBMs. We used CBM77 (binds to pectin, shown in green), CBM3a (binds to "semi-crystalline" cellulose, shown in cyan) and CBM1Cel6a (binds to crystalline cellulose, shown in red). 3D confocal fluorescence microscopy images of UPF and PF are shown in figure 7 a) and b).

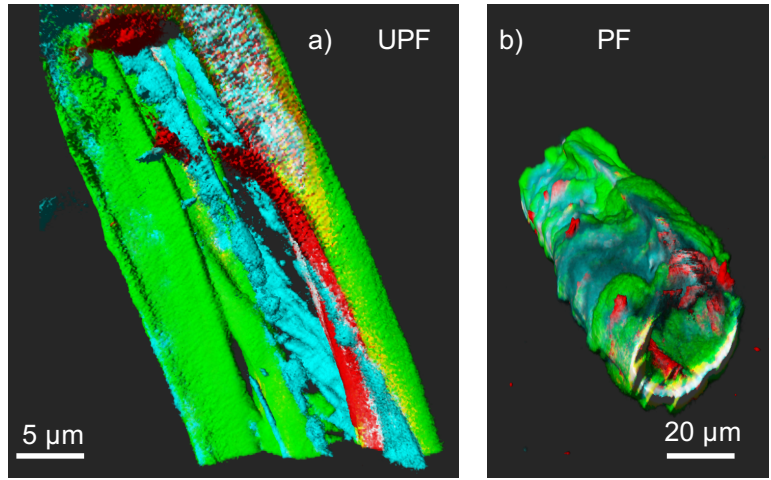

Figure 7: 3D images of confocal fluorescence microscopy imaging. The fibres are labelled with CBM77 in green (binds to pectin), with CBM3a in cyan (binds to "semi-crystalline" cellulose) and with CBM1Cel6a in red (binds to crystalline cellulose). a) UPF fibre and b) PF fibre.

Both fibre types exhibit a green (pectin) outer layer with cyan (semi-crystalline cellulose) parts and a red (crystalline cellulose) fibre interior (see figure 7). Intensity cross sections on single confocal planes reveal the order in which the three labelled components are arranged within a fibre wall (figure 8). By analysing the peak position of the labels, it is found that for UPF, the order is green, blue, green and red, i.e., pectin, semi-crystalline cellulose, pectin, and crystalline cellulose. The pectin signal is found on the outside and decays towards the fibre interior but is not found on the inside of the fibre wall. The signal for semi-crystalline cellulose, in turn, is found only in the fibre wall's interior, and the signal for crystalline cellulose is found only inside. In mature fibres, the P wall contains pectin, hemicelluloses, disordered cellulose fibres and lignin, and the S1 layer of the secondary cell wall has portions of pectin and hemicelluloses in addition to densely packed parallel cellulose microfibrils. Thus, both layers should occur in green but can also occur in blue, as both walls also contain "semicrystalline" cellulose. Therefore, it can be assumed that the green and blue intensity peaks represent the P wall and S1 layer. The crystalline cellulose is marked in red and can be attributed to the S2 layer of the secondary cell wall.

In contrast, in PF, the intensity peaks of the three components are closer together, but their signal distribution is broader. Similar to UPF, the order of the intensity peaks occurring towards the fibre interior is green, blue and red. However, in PF, only one peak is found for the green signal, which is consistent with the milling off of parts of P and S1 during the paper production process.

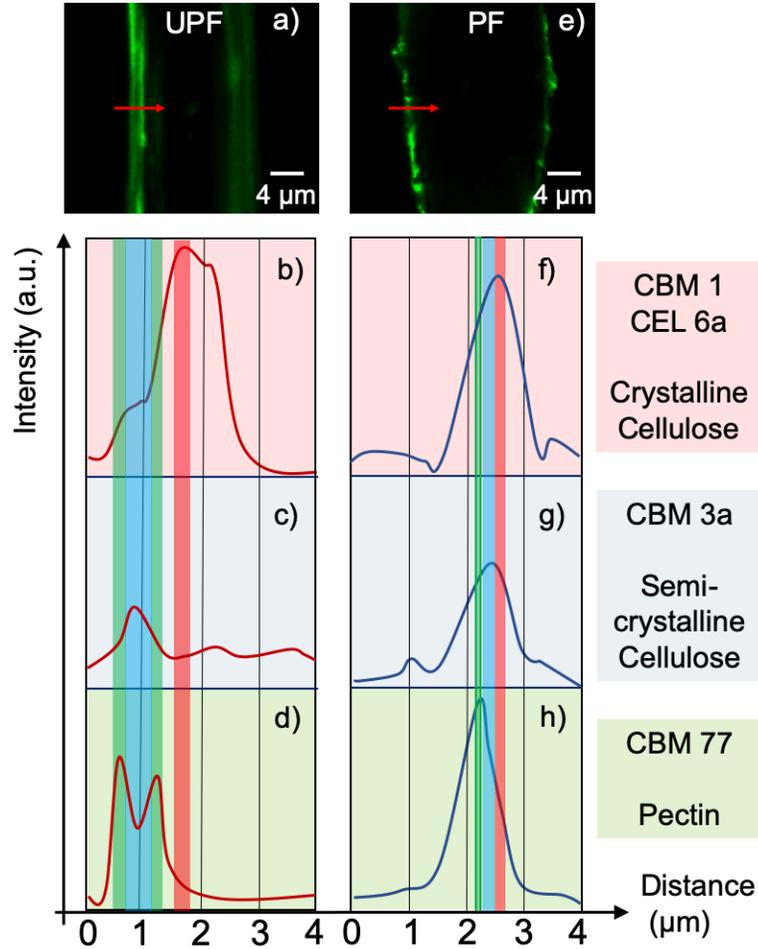

Figure 8: a) Image slice of the confocal fluorescence imaging of a UPF. The intensity peaks over the distance of the CMBs in the UPF are shown in b) for CBM1Cel6a, in c) for CBM3a and in d) for CBM77. In e), an image slice of the confocal fluorescence imaging of a PF is shown. The intensity peaks over the distance of the CMBs in the PF are shown in f) for CBM1Cel6a, in g) for CBM3a and in h) for CBM77. The peaks of the corresponding CBMs are marked in the green (CBM77), blue (CBM3a) and red (CBM1Cel6a) boxes.

It was not possible to label the C layer or the different layers P, S1, S2 individually with CBMs. The measured intensities and intensity peaks could therefore also display a mixture of the different layers of walls. Furthermore, the fluorescence CLSM images have a lateral resolution of 250 nm and an axial resolution of 500 nm. Thus, it is not possible to assign the $E_{lok}(z)$ measured with AFM directly to the different wall layers with confocal microscopy. However, the measured order of the occurring intensity peaks and the correlation of the labelled parts with the predicted fibre structure strongly supports our interpretation of the AFM measurements.

In the literature, tensile testing showed that an intact S2 layer is the most important factor for the mechanical properties of the fibres (Barnett and Bonham 2004; Spokevicius et al. 2007). From nanoindentation experiments, it was inferred that the elastic modulus of the S2 layer should be higher than the elastic modulus of the other walls (Gindl and Schoberl 2004). Merely, in the work of (Bergander and Salmen 2000), the transversal elastic modulus was not highly dependent on S2 but on S1 and S3. The presented results in our work indicate that the hardness of the walls/layers and the corresponding meaning of the mechanical properties must be divided in the transverse and longitudinal directions. In tensile tests, the longitudinal elastic modulus of the fibres is determined by pulling at both ends of the fibres. The fibril structure of the S2 layer in the fibre is therefore oriented in the tensile direction. Hence, a higher resistance in the tensile direction is achieved. In contrast to the tensile test, in the force-volume mapping method, the fibre is not pulled at either end, but the stiffness at depth is measured by nanoindentation. Thus, the loading direction is perpendicular to the orientation of the fibrils of the S2 layer (figure 9b). Due to the geometry of the indenters used in conventional nanoindentation experiments, it is not possible to obtain a lateral resolution, as it is by the AFM force-volume mapping method. The difference in the arrangement of the applied force in relation to the fibril orientation in the S2 layer in the fibre is shown in figure 9. Our results are in line with (Bergander and Salmen 2000), who showed that the transversal elastic modulus does not depend on S2. This leads to the interpretation that a cellulose fibre is resistant to tensile forces mainly due to the orientation of the fibril structure in the S2 layer in the tensile direction. The resistance against compression forces is due to a hard layer on top of the surface, i.e., the fibril structure in the P wall.

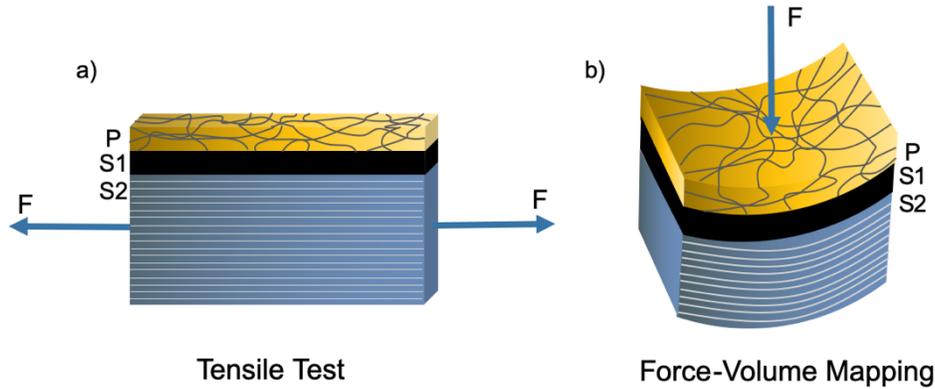

Figure 9: Fibril orientation in the applied load direction in a) the tensile test and b) force-volume mapping.

In the next experiment, PF and UPF were hydrated to investigate the difference in indentation depth and behaviour of the layered wall structures of the fibres.

In figure 10, the cross sections of $E_{lok}(z)$ in the xz-direction of a hydrated PF mapped with an HSC cantilever (a) and with a HAR cantilever (b) are shown. A cross section of a hydrated UPF is displayed in c) mapped with an HSC cantilever and in d) mapped with an HAR cantilever.

The hydrated PF exhibited fewer areas with a hard top layer on the fibre surface (yellow). In most parts, the PF showed areas that were attributed to a softer $E_{lok}(z)$ (blue or white). In these volume segments, the water molecules could break hydrogen bonds, which resulted in a softening of the fibre (Cabrera et al. 2011; Gumuskaya et al. 2003; John and Thomas 2008; Lindman et al. 2010). In only a few areas, the layered structure, observed in the dry state, was visible. Therefore, it is interpreted that in the hydrated state, the layered wall structure of the PF becomes indistinct. The bonding between the fibrils in each layer could be weakened due to the weakening of the H-bonds. Thus, a possible interpretation is that an intact wall structure in each layer cannot be sustained in the wet state. In the hydrated state of the UPF (figure 10c and d), a layered structure in the $E_{lok}(z)$ was observable in most parts of the cross section of the fibre. In contrast to the PF, the UPF seemed to preserve the layered wall structure in the hydrated state, which is suggested to be due to the intact C layer and P wall, which prevent the UPF from excessive softening in the hydrated state. The higher maximum value of $E_{lok}(z)$ in the UPF compared to the PF supports this consideration. However, the maximum value of $E_{lok}(z)$ in the hydrated UPF exhibits a lower value than in the dry state. As shown in figure 10, the indentation depth of a PF in the hydrated state is higher than that of the UPF.

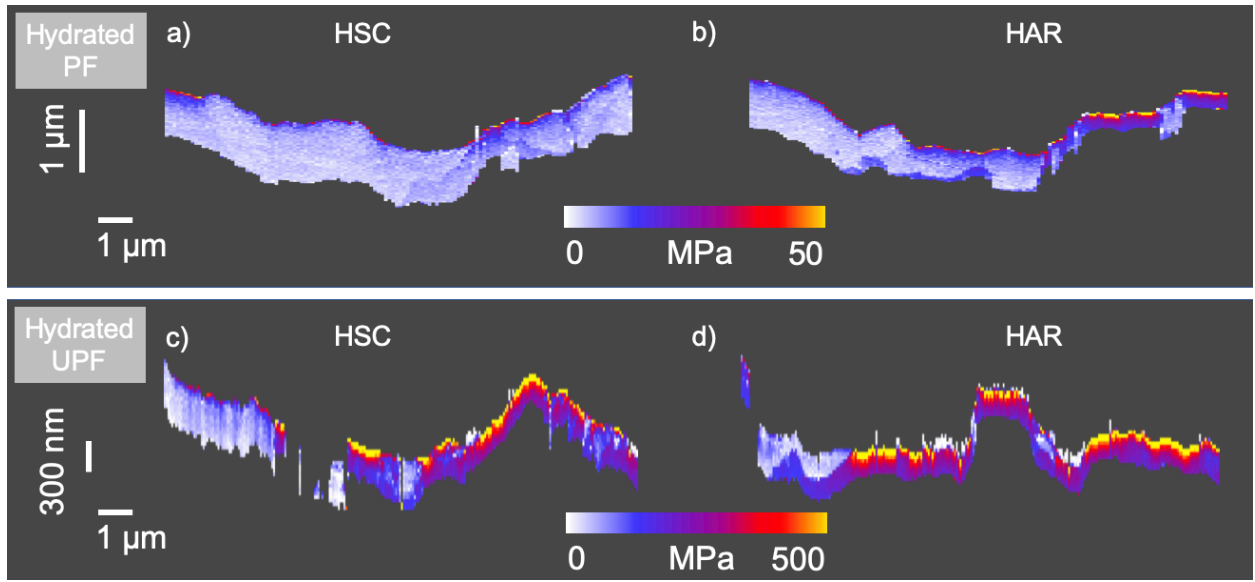

Figure 10: Cross section of $E_{lok}(z)$ in the xz-direction of a hydrated PF mapped with a) an HSC cantilever and b) an HAR cantilever. Cross section of $E_{lok}(z)$ in the xz-direction of a hydrated UPF mapped with c) an HSC cantilever and d) an HAR cantilever.

The indentation in the dry state of both fibres was normalised to 1. In both fibre types, a decay of the normalised indentation in the hydrated condition was observable. The decrease in the normalised indentation in the wet state compared to the dry state of the PF ranged from 1 to 0.16 ±0.14 and hence was larger than the decrease in the UPF to 0.81 ±0.05. Thus, the tip could indent into the PF deeper in the hydrated state than into the UPF. This is attributed to the fibre or pulp treatment before the paper-making process. As mentioned before, the C and small amounts of P are removed during the fibre or pulp treatment process in the PF. Therefore, predominantly cellulose-based layers are present. Hence, the normalised indentation in the PF is attributed to the greater softening of the entire cellulose-based fibre. In contrast, in the UPF, the wax-like layer C and the wax-, pectin- and lignin-containing P wall are still intact (Hartzell-Lawson and Hsieh 2000). With the intact composition and structure in the UPF, the water molecules cannot easily diffuse into the fibre. This can be seen as a natural hydrophobization to prevent massive water uptake into the UPF. Additionally, the standard deviation of the UPF was ±0.05 lower than that of the PF (±0.12). The more uniform behaviour is consistent with an intact layer structure. As all UPFs were extracted from a natural, raw cotton boll, and the fibres had the same unprocessed layer structure. Thus, these measurements were completed with reproduceable samples, unlike the processed PF. In the fibre or pulp treatment process, the PF pass processes, such as beating or pressing, where the original fibre structure is randomly destroyed in each fibre. Thus, the higher standard deviation in the PF originates from statistics in the fibre or pulp treatment before the paper-making process.

Furthermore, the influence of the cantilever tip system geometry was probed. An HSC and an HAR cantilever were used to investigate the indentation behaviour. The setpoint for both cantilevers was 3000 nN. A higher indentation depth was reached with the HSC cantilever. The difference between the two indentation depths was $15 \pm 4\%$. It was interpreted that with the HAR cantilever, the cantilever could not displace as much volume with its lower spring constant as the HSC cantilever. However, in all measurements, the choice of the cantilever was insignificant, as in all measurements, the desired indentation depth was reached. Figure S2 shows that the force-volume mapping method does not leave an imprint on the fibre surface in the scanned area with both types of cantilevers in the dry and wet states.

# Conclusion

We discuss an AFM-based method to assess the intrinsic mechanical properties of cellulosic fibres (processed linters and unprocessed cotton). With the local analysis of recorded static force-distance curves, we were able to measure the local elastic moduli at the fibre surface and 950 nm beneath the surface. This approach was used to investigate how fibre or pulp treatment before paper making affects the mechanical properties of cotton linter fibres at the surface.

In a combined 3D representation, topographic features could be directly related to the local mechanical properties. Dry fibres had a stiffer outer layer (higher $E_{lok}(z)$) on the fibre surface with softer material (lower $E_{lok}(z)$) beneath. The surface of UPF was stiffer than that of the processed fibres. For hydrated fibres, however, differences in the stiffness profile between the processed and unprocessed stiffnesses were found. The maximum indentation depth was much higher for the processed fibres than for the unprocessed fibres. This observation can be explained assuming that during the fibre or pulp treatment process, the C layer and perhaps the P wall were affected in the milling process. Thus, the processed fibres exhibited a lower $E_{lok}(z)$ layer on the fibre surface than the unprocessed fibres where the wall structure was intact. Assuming that intact wax-containing cuticle (C) and primary cell walls (P) serve as protection against intruding water molecules explains why the mechanical properties of the unprocessed fibre were less affected by hydration than those of the linters.

These findings are supplemented by microscopic observations. On the one hand, the data indicate that the P wall is missing in PF, as the external pectin signal is missing but the primary wall would have a high proportion of it. On the other hand, they show that the fibre wall in PF is less clearly layered than in UPF, as the signal intensity peaks are much closer together and at the same time broader than in UPF. Hence, processing not only removes the P wall but also affects the clear stratification and orderly structure of the fibre wall. Compaction, in turn, may result if a structurally challenged wall swells and is then dried during the paper-making process.

To investigate the influence of tip and cantilever properties on the results, data obtained with (relatively soft) cantilevers with high-aspect-ratio tips were compared to data obtained with a high spring constant cantilever with a standard tip. With both approaches, i.e., either using stiff cantilevers or using high aspect ratio tips, an indentation depth of a minimum of 300 nm could be achieved, and the mechanical stiffness was probed as a function of indentation depth. A higher maximum indentation depth was achieved using the hard cantilever HSC.

The subsurface imaging method proved to be a valuable tool for surface near depth-sensitive mapping of the local stiffness of cellulosic fibres. Local, i.e., lateral and vertical variations in the mechanical properties could be investigated and related to features of the fibre wall. Furthermore, the recovering behaviour of the layered wall structure from the hydrated to the dry condition and the potential fatigue could be investigated with the presented AFM-based method.

# Declarations

**Funding**

The work was supported by Deutsche Forschungsgemeinschaft under grant PAK 962, project numbers 405549611, 405422473, 405440040 and 52300548.

**Conflicts of interest/Competing interests**

The authors have no conflicts of interest to declare that are relevant to the content of this article.

**Availability of data and material**

Additional material is available in the supplementary information.

**Authors' contributions**

JA wrote the manuscript and conducted all AFM measurements and analysis. ML performed the confocal fluorescence microscopy imaging measurements. TK performed the SEM measurements. J-LS prepared the paper sheets and helped with knowledge about paper fibres. TM, MB and RWS planned and supervised the whole project and are responsible for any correspondence. All authors contributed to the writing of the manuscript.

**Ethics approval**

The manuscript is not submitted elsewhere and is an original research. There are no animal and human studies involved in this research. The authors have no conflicts of interest to declare.